\documentclass[letter,twocolumn]{jpsj3}

\usepackage{amsmath}
\usepackage{amsfonts}
\usepackage{amssymb}
\usepackage{txfonts}
\usepackage{bm}
\usepackage{tabularx}
\usepackage{graphicx,color}
\usepackage{cite}

\def\Vec#1{\bm{#1}}
\def\Hc2{H_\mathrm{c2}}
\def\Tc{T_\mathrm{c}}

\author{
	Shunichiro Kittaka$^1$\thanks{E-mail: kittaka@issp.u-tokyo.ac.jp}, 
	Yusei Shimizu$^{1}$,
	Toshiro Sakakibara$^{1}$, 
	Yoshinori Haga$^{2}$,
	Etsuji Yamamoto$^{2}$,
	Yoshichika \={O}nuki$^{2,3}$,
	Yasumasa Tsutsumi$^{4,5}$,
	Takuya Nomoto$^{6}$, 
	Hiroaki Ikeda$^{7}$, and
	Kazushige Machida$^{7}$
}
\inst{$^{1}$Institute for Solid State Physics, University of Tokyo, Kashiwa, Chiba 277-8581, Japan\\
      $^{2}$Advanced Science Research Center, Japan Atomic Energy Agency, Tokai, Ibaraki 319-1195, Japan\\
      $^{3}$Department of Physics, University of the Ryukyus, Nishihara, Okinawa 903-0213, Japan\\
      $^{4}$Department of Basic Science, University of Tokyo, Tokyo 153-8902, Japan\\
      $^{5}$Condensed Matter Theory Laboratory, RIKEN, Wako, Saitama 351-0198, Japan\\
      $^{6}$Department of Physics, Kyoto University, Kyoto 606-8502, Japan\\
      $^{7}$Department of Physics, Ritsumeikan University, Kusatsu, Shiga 525-8577, Japan
}
\begin{document}

\title{Evidence for Chiral $d$-Wave Superconductivity in URu$_2$Si$_2$ \\from the Field-Angle Variation of Its Specific Heat}

\date{\today}

\abst{
Low-energy quasiparticle (QP) excitations in the heavy-fermion superconductor URu$_2$Si$_2$ were investigated by specific-heat $C(T,H,\phi,\theta)$ measurements of a high-quality single crystal.
The occurrence of QP excitations due to the Doppler-shift effect was detected regardless of the field direction in $C(H)$ of the present clean sample, 
which is in sharp contrast to a previous report.
Furthermore, the polar-angle-dependent $C(\theta)$ measured under a rotating magnetic field within the $ac$ plane exhibits 
a shoulder-like anomaly at $\theta \sim 45^\circ$ and a sharp dip at $\theta=90^\circ$ ($H \parallel a$) in the moderate-field region.
These features are supported by theoretical analyses based on microscopic calculations assuming 
the gap symmetry of $k_z(k_x+ik_y)$, whose gap structure is characterized by a combination of a horizontal line node at the equator and point nodes at the poles.
The present results have settled the previous controversy over the gap structure of URu$_2$Si$_2$ and have authenticated its chiral $d$-wave superconductivity.
}

\maketitle
The pairing mechanism of unconventional superconductivity in the heavy-fermion compound URu$_2$Si$_2$ ($\Tc=1.4$~K) has been one of the most tantalizing questions in condensed-matter physics.
It coexists with the  ``hidden-order'' (HO) phase occurring below $T_{\rm HO}=17.5$~K, a mysterious phase that exhibits neither an ordinary magnetic order nor a structure change~\cite{Mydosh2011RMP}.
The true nature of the HO phase is a long-standing mystery and has been intensively studied by a number of experimental and theoretical groups.
High-pressure experiments have revealed that superconductivity disappears and the HO phase transforms to the antiferromagnetic phase with a large moment at a critical pressure of $\sim 0.7$~GPa~\cite{Amitsuka2007JMMM}. 
The Shubnikov-de Haas experiments performed under pressure~\cite{Hassinger2010PRL} indicate that the Fermi surface is similar between the HO and antiferromagnetic phases.
On the basis of a first-principles theoretical approach, emergent rank-5 nematic order has been proposed as a possible origin of the HO phase~\cite{Ikeda2012NatPhys},
which can naturally explain the rotational symmetry breaking suggested 
from recent experiments~\cite{Shibauchi2014PM,Okazaki2011Science,Tonegawa2012PRL,Tonegawa2014NatCom,Kambe2014PRL}.

Towards the identification of the pairing mechanism of superconductivity, the determination of the gap symmetry is essential 
because it is closely related to the pairing glue.
The spin part of the order parameter for URu$_2$Si$_2$ has been considered to be singlet because the Pauli-paramagnetic effect is observed in any field direction~\cite{Brison1995PhysicaC},
although a change in the NMR Knight shift below $\Tc$ has not yet been detected~\cite{Kohori1996JPSJ}.
Recent polar Kerr effect measurements~\cite{Schemm2015PRB} and other experiments~\cite{Li2013PRB,Kawasaki2014JPSJ} have provided evidence for
the spontaneous breaking of time-reversal symmetry in the superconducting state, i.e., a chiral state in which two order parameters are degenerated.
Indeed, a giant Nernst effect has recently been observed above $\Tc$~\cite{Yamashita2015NatPhys}, and has been interpreted by asymmetric scatterings due to chiral superconducting fluctuations~\cite{Sumiyoshi2014PRB}.
It has been widely accepted that the superconducting gap possesses nodes on the basis of the power-law temperature dependence of thermodynamic quantities~\cite{Fisher1990PhysicaB,Matsuda1996JPSJ,Kasahara2007PRL}. 
The invariance of the specific heat~\cite{Yano2008PRL} and the thermal conductivity~\cite{Kasahara2007PRL,Kasahara2009NJP} with changing in-plane field direction
demonstrates the absence of vertical line nodes.
From these experimental findings, chiral $d$-wave superconductivity, the $k_{z}(k_{x}+ik_{y})$ pairing state, has been inferred.
For this state, the superconducting gap is expected to have a horizontal line node at $k_z=0$ and point nodes at $k_x=k_y=0$.

Nevertheless, the detection of the detailed nodal structure of URu$_2$Si$_2$ remains controversial.
Previous measurements of the specific heat performed at 0.34~K~\cite{Yano2008PRL} indicated the presence of point nodes at the north and south poles  but did not detect the horizontal line node in the heavy-mass bands. 
This conclusion was derived from the observation that the specific heat showed $H$-linear (nearly $\sqrt{H}$-linear) dependence in the low-field region for $H \parallel c$ ($H \perp c$).
On the basis of the Doppler-shift effect~\cite{Volovik1993JETPL}, $\delta E \propto \Vec{v}_{\rm F}\cdot \Vec{v}_{\rm s}$, the results suggest that Doppler-shift-active quasiparticles (QPs) are absent for $H \parallel c$, i.e., all the nodal QPs have their $\Vec{v}_{\rm F}$  parallel to the $z$ direction.
Here, $\Vec{v}_{\rm F}$ is the Fermi velocity at a node and $\Vec{v}_{\rm s}$ is the supercurrent velocity circulating around each vortex ($\Vec{H} \perp \Vec{v}_{\rm s}$).
Meanwhile, thermal-conductivity studies~\cite{Kasahara2007PRL,Kasahara2009NJP} concluded a horizontal line node in the light-mass bands and point nodes in the heavy-mass bands.
Thus, experimental evidence for a horizontal line node in the heavy-mass bands has been missing.

Regarding the ordered-state Fermi surface of URu$_2$Si$_2$, recent experimental and theoretical studies indicate the presence of heavy-mass sheets around the M point~\cite{Ikeda2012NatPhys,Tonegawa2013PRB,Shibauchi2014PM}.
Therefore, the absence of a horizontal line node on the heavy-mass band is inconsistent with the anticipated chiral $d$-wave state.
In order to settle this controversy, in this study we measured the field-angle-dependent specific heat down to a low temperature of 0.2~K. 
We used a high-quality single-crystalline sample (10.6 mg weight) grown by the Czochralski pulling method in a tetra-arc furnace~\cite{Matsuda2011JPSJ}.
The residual resistivity ratio of another crystal from the same batch is 255.
The specific heat $C$ was measured by the standard quasi-adiabatic heat-pulse and relaxation methods in a dilution refrigerator.
The magnetic-field orientation was controlled with high accuracy using a vector magnet.

\begin{figure}
\begin{center}
\vspace{0.1in}
\includegraphics[width=2.9in]{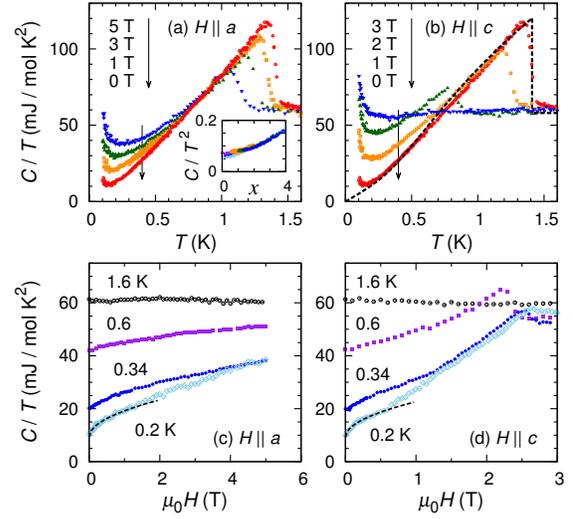}
\caption{
(Color online) 
Temperature dependence of the specific heat divided by temperature, $C/T$, measured at several magnetic fields for (a) $H \parallel a$ and (b) $H \parallel c$. 
Field dependence of $C/T$ measured at 0.2, 0.34, 0.6, and 1.6~K for (c) $H \parallel a$ and (d) $H \parallel c$.
In the inset of (a), $C/T^2$ in the selected $T$ and $H$ regions for $H \parallel a$ is plotted as a function of $x=\sqrt{(H/\Hc2)}/(T/\Tc)$ in the unit of J/(mol K$^3$).
The dashed line in (b) is obtained by calculation assuming the gap symmetry of $k_z(k_x+ik_y)$ with a gap size of $\Delta_0=1.6k_{\rm B}\Tc$.
The dashed lines in (c) and (d) are fits to the data by using the function $C(H)=a\sqrt{H}+C(0)$.
}
\label{CTH}
\vspace{-0.3in}
\end{center}
\end{figure}

Figures \ref{CTH}(a) and \ref{CTH}(b) show the temperature dependence of $C/T$ measured under a magnetic field parallel to the $a$ and $c$ axes, respectively.
The zero-field $C/T$ shows a sharp jump at $\Tc$ of 1.41 K and takes relatively small values at low temperatures below 0.2~K.
These results ensure that the present sample is of higher quality than the sample used in the previous study~\cite{Yano2008PRL}. 

In the intermediate-$T$ region between 0.3 and 1~K, the zero-field $C/T$ exhibits a nearly $T$-linear dependence as previously observed.
The extrapolation of this $T$-linear behavior to $T=0$, which yielded a positive intersect in previous reports~\cite{Fisher1990PhysicaB, Matsuda2011JPSJ, Yano2008PRL}, becomes negative for the present clean sample.
Apparently, a simple $C(T) \propto T^2$ dependence does not hold in the whole temperature range below 1~K and the slope of $C(T)/T$ becomes intrinsically smaller below 0.3~K.
These features, along with a small jump of the specific heat at $\Tc$, $\Delta C/\gamma\Tc \sim 1$, 
can be reproduced approximately from a single-band calculation by assuming the gap symmetry of $k_z(k_x+ik_y)$ 
with a gap size of $\Delta_0=1.6k_{\rm B}\Tc$, as represented by the dashed line in Fig.~\ref{CTH}(b).
Incorporating the multiband effect would improve the agreement between the calculated result and the experimental data.

The slight upturn in $C/T$ on cooling below 0.2~K is probably due to some impurity effect, 
because it becomes larger in lower-quality samples~\cite{Matsuda2011JPSJ}.
With increasing the magnetic field, this upturn significantly develops in any field direction.
The field-enhanced upturn in $C/T$ is too large to be explained by the nuclear Schottky contribution.
To exclude the effect of this unusual upturn, we hereafter concentrate on the data measured above 0.2~K.

For $H \parallel c$, the sample is always in the normal state at 3~T.
As shown in Fig.~\ref{CTH}(b), the normal-state $C/T$ decreases slightly on cooling to 0.2~K.
Similar behavior has also been observed in previous results~\cite{Matsuda2011JPSJ}.

Figures \ref{CTH}(c) and \ref{CTH}(d) represent the field dependence of $C/T$ at 0.2, 0.34, 0.6, and 1.6~K for $H \parallel a$ and $H \parallel c$, respectively.
The $C(H)$ data measured at 1.6~K show a slight change with $H$, reflecting its normal-state property.
Because of the limited field range of the present experiment, the upper critical field  ($\mu_0\Hc2^{\parallel a} \sim 12$~T)  was not reached for $H \parallel a$.

Let us first focus on the $C(H)$ data at 0.34~K, taken under the same conditions as those described in Ref.~\ref{Yano2008PRL}. 
In sharp contrast to the previous report, in which no $\sqrt{H}$ behavior was observed for $H \parallel c$,
$C(H)$ for the present sample shows $\sqrt{H}$-like behavior not only for $H \parallel a$ but also for $H \parallel c$ in the low-field region. 
This change probably occurred owing to the improvement of the sample quality;
in general, the low-energy QP excitations are obscured 
easily by the impurity-scattering effect~\cite{Kubert1998SSC} and thermal QP excitations.
Indeed, by increasing the temperature to 0.6~K, the $\sqrt{H}$ behavior becomes indiscernible, particularly for $H \parallel c$.

To more clearly see the low-energy QP excitations in URu$_2$Si$_2$, we measured $C(H)$ at a lower temperature of 0.2~K.
The rapid increase in $C(H)$ at low fields is more clearly detected for both field directions.
These low-field data can be fitted satisfactorily by using the function $C(H)=a\sqrt{H}+C(0)$ as demonstrated by dashed lines in Figs.~\ref{CTH}(c) and \ref{CTH}(d).
Furthermore, we have confirmed that the present $C(T)$ and $C(H)$ data measured for $0.2~{\rm K} \le T \le 0.7~{\rm K}$ in $\mu_0H_{\parallel a} \le 5$~T obey 
the scaling law~\cite{Volovik1997PRL} $C(T,H)/T^2=F(x)$ in a wide $x$ region, 
as shown in the inset of Fig.~\ref{CTH}(a).
Here, $x=\sqrt{(H/\Hc2)}/(T/\Tc)$ and $F(x)$ is a scaling function that is constant (proportional to $x$) for small (large) $x$.
From these results, it is concluded that the heavy-mass bands have node(s) in regions where $v_{\rm F} \nparallel z$.
In other words, even the existence of point nodes at poles is not justified by the $C(H)$ data alone.

\begin{figure}
\begin{center}
\vspace{0.1in}
\includegraphics[width=2.9in]{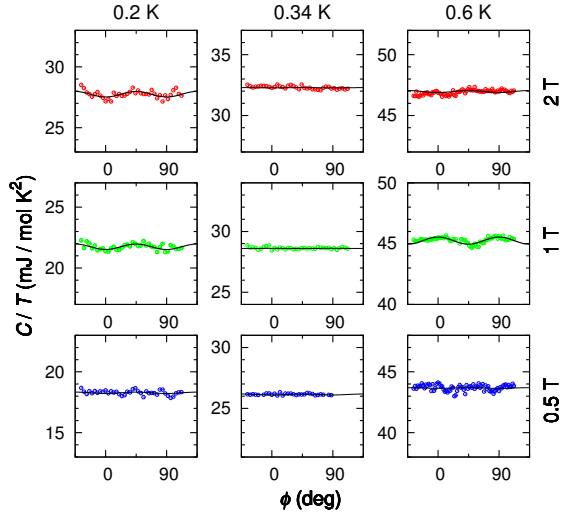}
\caption{
(Color online) 
Azimuthal-angle $\phi$ dependence of $C/T$ measured at 0.2, 0.34, and 0.6 K under selected magnetic fields rotated within the $ab$ plane. 
The field angle $\phi$ is measured from the $a$ axis.
Solid lines are fits to the data by $A_4(T,H)\cos4\phi+C_0(H)$. 
}
\label{Cphi}
\vspace{-0.3in}
\end{center}
\end{figure}

Then, we need to reexamine the gap structure of URu$_2$Si$_2$.
Figure $\ref{Cphi}$ shows the azimuthal-angle $\phi$ dependence of the specific heat measured by rotating a magnetic field within the $ab$ plane at various field strengths and temperatures.
Here, the field angle $\phi$ is measured from the $a$ axis.
Recall that, according to Doppler-shift analyses~\cite{Vekhter1999PRB,Sakakibara2007JPSJ}, 
the total QP density of states shows a local minimum when the magnetic field is along the nodal or gap-minimum direction. 
At $T=0.34$~K, the specific heat is invariant with $\phi$ at any field, consistent with the previous report~\cite{Yano2008PRL}.
By contrast, at 0.2 and 0.6~K, we observed a clear fourfold oscillation in $C(\phi)$ at 1~T ($\sim 0.07\Hc2$).
Moreover, the sign of the oscillation changes at around 0.34~K, reminiscent of the $C(\phi)$ behavior in CeCoIn$_5$~\cite{An2010PRL}.
However, this fourfold oscillation is diminished by decreasing the magnetic field to 0.5~T ($\sim 0.035\Hc2$).
This behavior, which is similar to the case of the anisotropic $s$-wave superconductor CeRu$_2$~\cite{Kittaka2013JPSJ}, 
can be explained by the presence of gap minima and/or anisotropy of the Fermi velocity~\cite{Miranovic2003PRB,Miranovic2005JPC}. 
Thus, we conclude that vertical line nodes for the $\phi$ rotation are absent.

\begin{figure}
\begin{center}
\vspace{0.1in}
\includegraphics[width=2.9in]{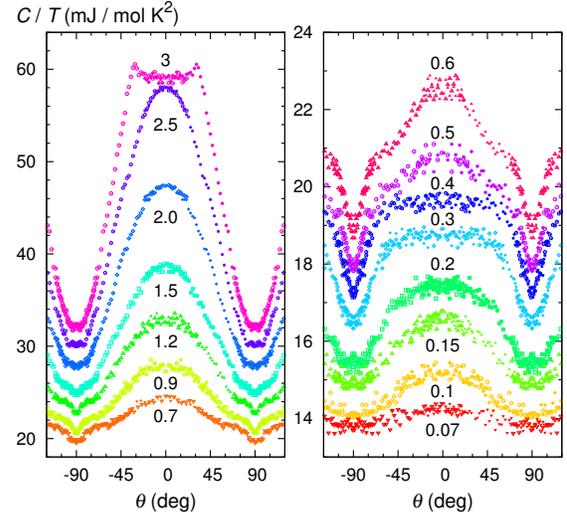}
\caption{
(Color online) 
Polar-angle $\theta$ dependence of $C/T$ measured at 0.2~K under various magnetic fields rotated within the $ac$ plane. 
The field angle $\theta$ is measured from the $c$ axis.
For clarity, the data measured in the interval $-20^\circ \lesssim \theta \lesssim 110^\circ$ (solid symbols) are plotted repeatedly (open symbols).
Numbers labeling the curves represent the magnetic field in tesla.}
\label{Ct}
\vspace{-0.3in}
\end{center}
\end{figure}

Next, to study the horizontal line nodes and polar point nodes,
we investigated the polar-angle $\theta$ dependence of the specific heat by rotating a magnetic field within the $ac$ plane.
Here, $\theta$ is the angle between the magnetic field and the $c$ axis.
The $C(H,\theta)$ data measured at 0.2~K are shown in Fig.~\ref{Ct}.
Note that the sample is in the normal state near $\theta = 0^\circ$ at 3~T because $\mu_0\Hc2^{\parallel c} \sim 2.6$~T [see Fig.~\ref{CTH}(d)].
In the high-field region, a large twofold $C(\theta)$ oscillation, reflecting the anisotropy of $\Hc2$ that is predominantly due to the anisotropic Pauli-paramagnetic effect, is observed.

In the low-field region below 0.15~T, no significant anomaly was detected in $C(\theta)$ within the sensitivity  limits of our measurements: 
the data can be fitted by a simple twofold function, $C(\theta,H)=A_2(H) \cos2\theta + C_0(H)$.
By contrast, a striking feature is found in $C(\theta)$ at 0.2~T; a shoulder-like anomaly appears at around $\theta \sim 45^\circ$. 
With increasing the magnetic field, this anomaly slightly moves to the higher $\theta$ side but is visible up to at least 1.5~T. 
In addition, the  dip at $\theta=90^\circ$ becomes sharp and pronounced in the intermediate-field region between 0.3 and 1.5~T.

\begin{figure}
\begin{center}
\vspace{0.1in}
\includegraphics[width=2.75in]{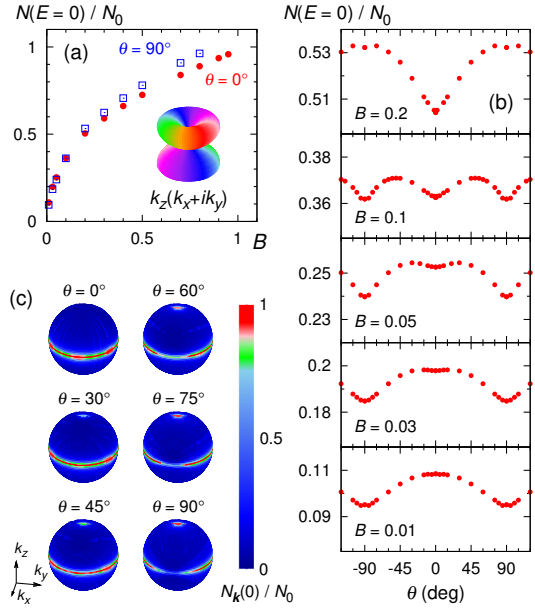}
\caption{
(Color online) 
(a) Field dependence of the zero-energy density of states $N(E=0)$ at $\theta=0^\circ$ ($H \parallel z$, circles) and $90^\circ$ ($H \parallel x$, squares),
obtained by microscopic calculations with the assumption of the gap symmetry of $k_z(k_x+ik_y)$ (inset).
Here, $N_0$ is the density of states in the normal state and $B$ is scaled by the Eilenberger unit~\cite{Miranovic2003PRB}.
(b) Polar-angle $\theta$ dependence of $N(E=0)$ at several fields.
(c) Angle-resolved density of states $N_{\Vec{k}}(E=0)$ at $B=0.03$ for selected $\theta$, mapped on the spherical Fermi surface.
Here, the magnetic field is rotated within the $xz$ plane.
}
\label{calc}
\vspace{-0.3in}
\end{center}
\end{figure}

In order to examine whether these features of $C(\theta)$ can be explained by the anticipated chiral $d$-wave state, 
we calculate the polar-angle dependence of the zero-energy density of states, $N(E=0)$, 
on the basis of the microscopic Eilenberger theory assuming the gap symmetry of $k_z(k_x+ik_y)$ [inset of Fig.~\ref{calc}(a)] and a single-band spherical Fermi surface~\cite{Tsutsumi2015}.
For simplicity, the Pauli-paramagnetic effect and the anisotropy of the Fermi velocity are not taken into account.
Figure \ref{calc}(a) shows the field dependence of $N(E=0)$ at $\theta=0^\circ$ ($H \parallel z$) and $90^\circ$ ($H \parallel x$).
Both data show $\sqrt{H}$ behavior, as observed in the present experiment.
At low fields, $N(E=0)$ is slightly larger for $H \parallel z$ than for $H \parallel x$.
By increasing the magnetic field, this anisotropy is reversed at $B\sim 0.1$ 
because of the slight anisotropy of $\Hc2$ ($\Hc2^{\parallel x} \lesssim \Hc2^{\parallel z}$) originating from the gap structure.

In Fig. \ref{calc}(b), $N(E=0)$ is plotted as a function of the polar angle $\theta$ at several fields.
The calculated result at a low field of $B=0.03$ can reproduce the features observed in the present experiment, 
i.e., the shoulder-like anomaly and the dip structure in $C(\theta)$ observed at moderate fields.
With increasing the magnetic field, a sharp dip develops at $\theta=0^\circ$ and then the anisotropy in $N(E=0)$ is reversed above $B \sim 0.1$.
This reflects the $\Hc2$ anisotropy, as mentioned above.
The dip at $\theta=0^\circ$ and the reversal of the $C(\theta)$ anisotropy in the high-field region,
demonstrated in Fig.~\ref{calc}(b), were not detected in the experiment 
because of the actual $\Hc2$ anisotropy $\Hc2^{\parallel a} > \Hc2^{\parallel c}$ owing to the strong Pauli-paramagnetic effect.

To determine the reason why the shoulder-like anomaly appears, 
the ${\Vec{k}}$-resolved density of states, $N_{\Vec{k}}(0)$, obtained from the present calculations is mapped on the spherical Fermi surface in Fig.~\ref{calc}(c).
Here, the magnetic field of $B=0.03$ is applied within the $xz$ plane. 
Note that the integration of $N_{\Vec{k}}(0)$ over the Fermi surface yields $N(E=0)$.
When the magnetic field is applied along the $z$ axis ($\theta=0^\circ$), 
prominent QP excitations occur anywhere near the horizontal line node while those at the point nodes are inactive.
By tilting the magnetic field away from the $z$ axis, QPs are also excited around the point nodes.

An important change in $N_{\Vec{k}}(0)$ can be seen between $\theta=45^\circ$ and $60^\circ$;
for $\theta \ge 60^\circ$, the contribution from the part of the horizontal line node near $k_y=0$ is strongly reduced
because the Doppler-shift effect becomes small there.
As a result, a substantial suppression of the total QP density of states occurs for $\theta \gtrsim 60^\circ$ and leads to the shoulder-like anomaly around $\theta \sim 45^\circ$ at $B=0.03$.
This $\theta$ dependence is less sensitive to the change in the Fermi-surface shape; 
qualitatively the same results can be obtained from calculations by using cigar-shaped and spheroidal Fermi surfaces.~\cite{Tsutsumi2015} 
Hence, the observed shoulder-like anomaly in $C(\theta)$ can be attributed to the existence of a horizontal line node at $k_z=0$.
Note that, in the even-parity basis functions for tetragonal-lattice symmetry, 
the presence of a horizontal line node at $k_z=0$ and the absence of vertical line nodes only match with the chiral $E_g$ symmetry~\cite{Sigrist1991RMP}.
Therefore, the present results strongly indicate 
that the gap symmetry of URu$_2$Si$_2$ is of the $k_z(k_x+ik_y)$ type.\cite{note1}

In summary, 
the field-orientation-dependent specific heat of URu$_2$Si$_2$ was measured by using a high-quality single crystal.
$\sqrt{H}$ behavior of the specific heat was observed 
in any field direction for the present clean sample, whereas it was not observed for $H \parallel c$ in the previous study~\cite{Yano2008PRL}.
A fourfold oscillation of the specific heat was found in a rotating magnetic field within the $ab$ plane, 
but it disappeared by lowering the magnetic field. 
This fact ensures the absence of vertical line nodes in the superconducting gap.
In addition, we found a shoulder-like anomaly at $\theta \sim 45^\circ$ and a sharp dip at $\theta=90^\circ$ in the polar-angle $\theta$ dependence of the specific heat.
From theoretical analyses, these features can be explained by assuming the presence of a horizontal line node at the equator and point nodes at the poles in the gap.
Thus, the gap symmetry of URu$_2$Si$_2$ has been identified as a chiral $d$-wave type described by $k_z(k_x+ik_y)$. 

\acknowledgments
We thank Y. Yanase for useful discussions.
A part of the numerical calculations was performed by using the HOKUSAI GreatWave supercomputer system in RIKEN.
This work was supported by a Grant-in-Aid for Scientific Research on Innovative Areas ``J-Physics'' (15H05883)
from MEXT, and KAKENHI (15K05158, 15H03682, 25103716, 26400360, 15H05745, 15K17715, 15J05698) from JSPS.

\bibliographystyle{jpsj2}
\bibliography{C:/usr/local/share/texmf/bibref/ref_URu2Si2.bib}

\end{document}